\documentclass[fleqn,aps,pra,amsmath,amssymb,preprint]{revtex4-1}

\usepackage{float}
\usepackage{graphicx}
\usepackage{textcomp}
\usepackage{braket}
\usepackage{enumerate}
\usepackage{multirow}
\usepackage{array}
\usepackage{tabularx}

\begin{document}

\title{Generation of a cold pulsed beam of Rb atoms by transfer from a 3D magneto-optic trap}
 
 \author{Sapam Ranjita Chanu}
 \affiliation{Department of Physics, Indian Institute of
 Science, Bangalore 560\,012, India}

 \author{Ketan D. Rathod}
 \affiliation{Department of Physics, Indian Institute of
 Science, Bangalore 560\,012, India}
 
\author{Vasant Natarajan}
\affiliation{Department of Physics, Indian Institute of
 Science, Bangalore 560\,012, India}
 \email{vasant@physics.iisc.ernet.in}
 \homepage{www.physics.iisc.ernet.in/~vasant}

\begin{abstract}
We demonstrate a technique for producing a cold pulsed beam of atoms by transferring a cloud of atoms trapped in a three dimensional magneto-optic trap (MOT). The MOT is loaded by heating a getter source of Rb atoms. We show that it is advantageous to transfer with two beams (with a small angle between them) compared to a single beam, because the atoms stop interacting with the beams in the two-beam technique, which results in a Gaussian velocity distribution. The atoms are further cooled in optical molasses by turning off the MOT magnetic field before the transfer beams are turned on. \\

\end{abstract}

\maketitle

\section{Introduction}

The generation of a slow atomic beam having a narrow velocity spread---longitudinally cold---has wide-ranging applications in fields such as atomic clocks \cite{YSM01,LBY15}, atom interferometry \cite{BER97}, and precision measurements \cite{WYC93}. In addition, one of the key requirements for experiments in quantum information processing \cite{RZC05} and quantum metrology \cite{BFM11} is the ability to prepare atoms in a particular quantum state. Both these requirements are met by having a cold beam of atoms transferred from a trap.

A variety of techniques have been used in earlier work to generate such cold atomic beams. In one well-established technique, atoms are first cooled and trapped in a magneto-optic trap (MOT), and then transfered. The MOT can be either two dimensional (2D) or three dimensional (3D). Atoms can be loaded into the MOT using either the low-velocity tail of background vapor (vapor MOT) \cite{DSW98,BJD98,CRU06,RWN01}, or using a Zeeman slower \cite{YKK03,SVH05,PRS10}. In fact, cold atomic beams can be directly generated using the force from counter-propagating photons in both Zeeman slowing schemes \cite{PHM82} and using frequency-chirped lasers \cite{EBH85}. However, these methods have the problem of the cooling laser beam being present along with the atomic beam (as discussed in Ref.\ \cite{RSN13}). A similar problem exists with the use of two laser beams in 1D moving-molasses configuration \cite{CSG91}. We have solved this problem by having inclined beams to generate the moving molasses beams \cite{RSN13}, and were then able to generate a continuous atomic beam. However, the longitudinal temperature was about 125 mK, which is significantly higher than the temperature obtained here. Cold atomic beams have also been generated by having a hole in the apex of a pyramidal MOT \cite{AMW98,CPC01}, but this requires non-standard beam polarizations and hence is not widely used since it was invented.

In this work, we report generation of a cold \textit{pulsed} beam of $^{87}$Rb atoms transfered from a 3D MOT. The atoms are transfered from the MOT using two different configurations of pushing beams---one with a single beam and the second using a pair of inclined beams. Atoms are loaded into the MOT using the relatively easy technique of heating a getter source, as demonstrated by us in earlier work \cite{RWN01}. The trap is formed by a standard configuration consisting of six intersecting laser beams and a quadrupole magnetic field generated using a pair of anti-Helmholtz coils. The atoms are allowed to thermalize to a lower temperature in optical molasses \cite{LPR89} after the MOT magnetic field is turned off. The atoms are then transfered from the MOT by turning on the pushing beam(s).

The velocity distribution of the transferred cloud is determined by time-of-flight at a distance of 330 mm from the trap center. MOT transfer with a single beam shows a non-Gaussian distribution with a long tail, which is to be expected since the atoms are always interacting with the pushing beam. On the other hand, the two-beam configuration results in a symmetric Gaussian distribution because the atoms stop interacting with the pushing beams once they are transferred. In the two-beam configuration the mean velocity is 27.5 m/s and the lowest temperature obtained is 6.4 mK. Two significant advantages of both kinds of MOT transfer are that -- (i) the transferred atoms are free from contamination from background atoms, and (ii) the atoms can be prepared in any desired $ F $ state.

\section{Experimental details}

\begin{figure}
\centering
\includegraphics[width=.5\columnwidth]{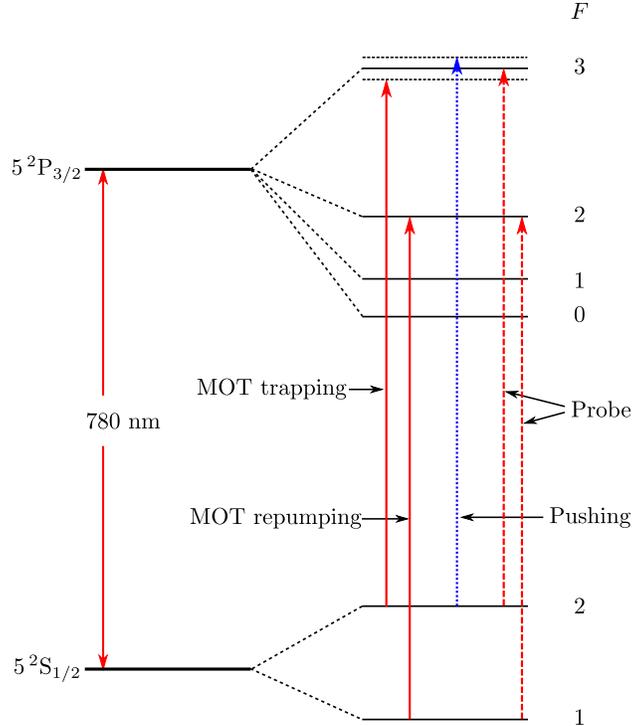}
\caption{(Color online) Low-lying energy levels of $^{87}$Rb -- not to scale. Hyperfine transitions (along with their $ F $ values) used in the experiment are shown.}
\label{fig:Rbmotenergy}
\end{figure}

The relevant energy levels of $^{87} \rm Rb $ are shown in Fig.\ \ref{fig:Rbmotenergy}. The transitions used for trapping and repumping in the MOT are indicated. The cold cloud is transferred with pushing beams at the wavelength shown, and the transferred atoms are probed using a resonant probe beam at the detection point.

All the laser beams are derived from home-built grating-stabilized diode laser systems. The frequency of the lasers are locked to appropriate peaks using saturated absorption spectroscopy in a vapor cell. The laser beams for trapping are generated from a master-slave laser configuration. Each trapping beam has a $1/e^2 $ diameter of 20 mm with power of 12 mW, and is detuned by 12 MHz ($2 \, \Gamma $) below the $ F_g = 2 \rightarrow F_e = 3 $ cooling transition. A repumping beam (derived from another laser) is required to prevent loss of atoms from the cooling transition. It is mixed with each trapping beam, and has the same size but with slightly smaller power. It is locked to the $ F_g = 1 \rightarrow F_e = 2 $ transition.

The pushing beam is slightly blue detuned from the cooling transition, so that it is resonant in a moving frame. The required frequency shift for this is produced using two acousto-optic modulators (AOMs). The transferred atoms are probed at a distance of 330 mm from the trap center by using a beam resonant with the cooling transition. It is linearly polarized, has a size of 3 mm, and a power of 2.5 mW. The probe beam also has a small amount of repumping light mixed into it. The fluorescence from the atoms is detected using a photomultiplier tube (Hamamatsu R928).

\begin{figure}
\centering
\includegraphics[width=.6\columnwidth]{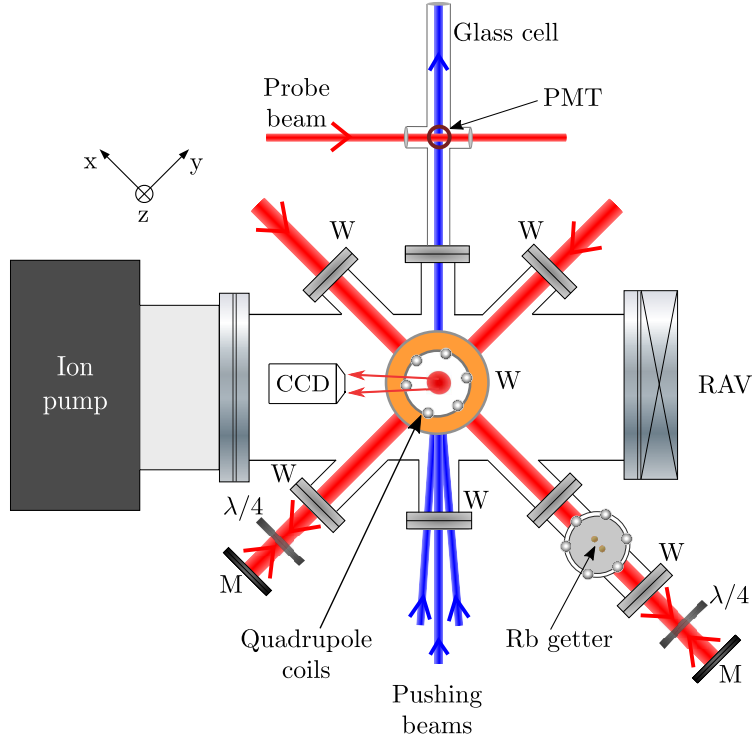}
\caption{(Color online)  Schematic of the vacuum chamber (top view) used in the experiment. Transfered atoms from the MOT are detected using fluorescence from a resonant probe beam. Figure key: W -- window; M -- mirror; $\lambda/4$ -- quarter waveplate; RAV -- right-angle valve; PMT -- photomultiplier tube.}
\label{fig:motchamber}
\end{figure}

The experimental vacuum chamber is shown schematically in Fig.\ \ref{fig:motchamber}. Optical access for the MOT beams is provided by glass viewports. The transferred atoms are probed in a glass cell. The MOT is loaded from the low-velocity tail of background vapor. The vapor itself is generated by heating a Rb getter source, as described in our earlier work in Ref.\ \cite{RWN01}. The getter source was mounted at a distance of 190 mm from the MOT center, as shown in the figure.

The MOT was formed at the intersection of 6 laser beams, obtained by retro-reflecting 3 beams in the 3 orthogonal directions. The beams were made to have circular polarization using quarter waveplates. The required quadrupole field was generated using a pair of coils mounted on the outside of the vacuum chamber. The MOT loading time was 4 s, determined by when the supply rate from the getter source is equal to the loss rate at the pressure of $ 3 \times  10^{-9} $ torr \cite{RWN01}. The trapped atoms were imaged with a CCD camera---a typical image shows an elliptic cloud of $10^7$ atoms when the field gradient is 10 G/cm. The size of the cloud is 4.1 mm $\times$ 1.4 mm. It is at a temperature of 300 \textmu K, which is the typical temperature in a MOT with cooling beams detuned by $-2 \, \Gamma$.

\begin{figure}
	\centering
	\includegraphics[width=.35\columnwidth]{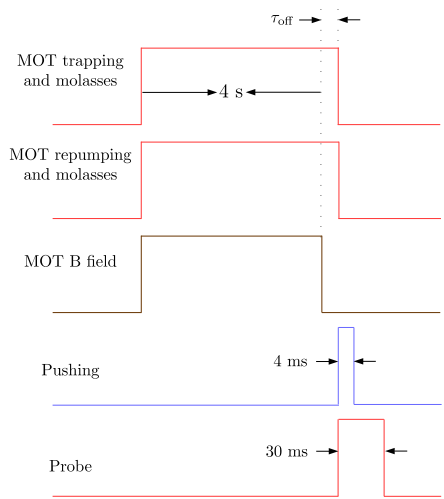}
	\caption{(Color online) Pulse sequence showing the timing of various beams and the MOT magnetic field turned on and off during the experiment (not to scale).}
	\label{fig:pulse_sequence}
\end{figure}

%
The atomic cloud from the MOT was transferred by switching on one or two pushing beams. The beams had a power of 2 mW and size of 3 mm. The beams were switched on using the same AOMs used for producing the frequency shift. The magnetic field of the MOT was turned off for a variable time $\tau_{\rm off}$, and the atoms are allowed to further cool in optical molasses before the transfer beams are turned on. Since the size of the cloud compresses in the molasses, the size of the pushing beam is much larger. The required timing, shown in Fig.\ \ref{fig:pulse_sequence}, was precisely controlled using a computer.

\section{Results and discussion}
\subsection{Transfer with one beam}

\begin{figure}
\centering
\includegraphics[width=.5\columnwidth]{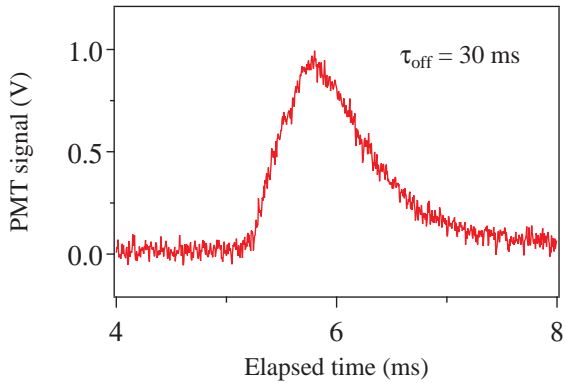}
\caption{(Color online)  Fluorescence signal as measured by the PMT for transfer with one beam. Time spent in the molasses is 30 ms.}
\label{fig:onebeam}
\end{figure}

These experiments are done with a single beam that is blue detuned from resonance by $\delta_s = + 7.4$ MHz. The detuning is chosen so that the transferred atoms reach the probe region within a few ms. A typical fluorescence spectrum measured by the PMT as a function of time elapsed from turning on the pushing beam is shown in Fig.\ \ref{fig:onebeam}. The thermalization time is 30 ms. The nature of the signal shows that the velocity distribution of the atoms is asymmetric, and is distorted from the initial Gaussian distribution in the MOT. This is because the atoms get accelerated as they continue to interact with the pushing beam all the way till they reach the probe beam. As we will see below, one of the main advantages of pushing with two beams is that the velocity distribution remains Gaussian.

\subsection{Transfer with two beams}

\begin{figure}
\centering
\includegraphics[width=.5\columnwidth]{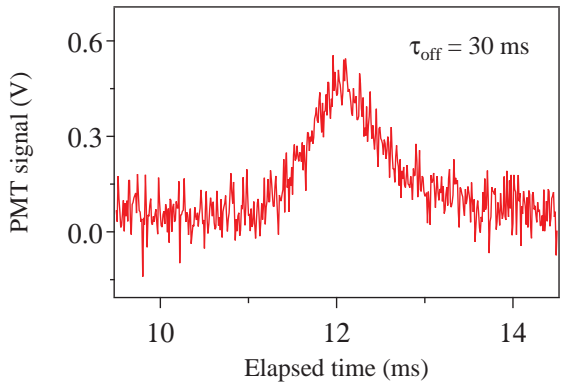}
\caption{(Color online)  Fluorescence signal as measured by the PMT for transfer by two beams. Time spent in the molasses is 30 ms.}
\label{fig:twobeamt}
\end{figure}

These experiments are done with two pushing beams with an angle of $18^\circ$ between them---both equally inclined from the vertical. The beams are detuned from resonance by $\delta_t = +7.05$ MHz, with the value again chosen so that the transfer occurs in a reasonable time. A typical fluorescence spectrum as a function of time elapsed from turning on the two transferring beams is shown in Fig.\ \ref{fig:twobeamt}. The symmetric nature of the spectrum is because the atoms pick up a terminal velocity once they move out of the region where they overlap with the transferring beams. The value of the terminal velocity depends on experimental parameters of detuning (which in turn determines the scattering rate), and the beam size and intersection angle (which determines the extent of overlap with the atoms in the MOT). For our conditions, the peak in the spectrum shows that the velocity is $0.33$ m / 0.012 s $= 27.5$ m/s.

\begin{figure}
\centering
\includegraphics[width=.5\columnwidth]{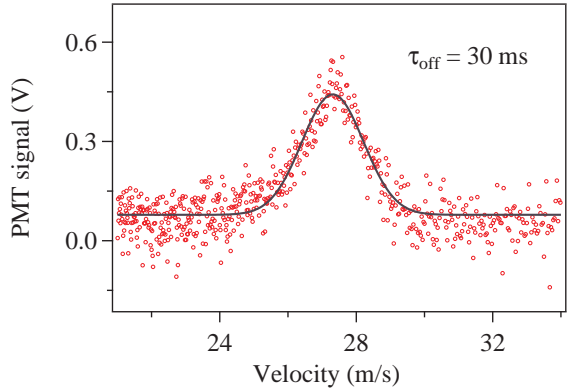}
\caption{(Color online)  Velocity distribution for transfer by two beams obtained from the fluorescence spectrum shown in Fig.\ \ref{fig:twobeamt}.}
\label{fig:twobeamvel}
\end{figure}

The symmetric nature of the spectrum allows us to convert the time axis to a velocity axis, simply using the fact the velocity at the probe point is the distance 0.33 m divided by the time it takes to get there. The results of this transformation for the above spectrum are shown in Fig.\ \ref{fig:twobeamvel}. The velocity distribution has a symmetric Gaussian lineshape as seen from the fit to the data, and is, as mentioned before, one of the primary advantages of the two-beam technique. The rms value from the Gaussian fit yields the temperature of the cloud.

\begin{figure}
\centering
\includegraphics[width=.5\columnwidth]{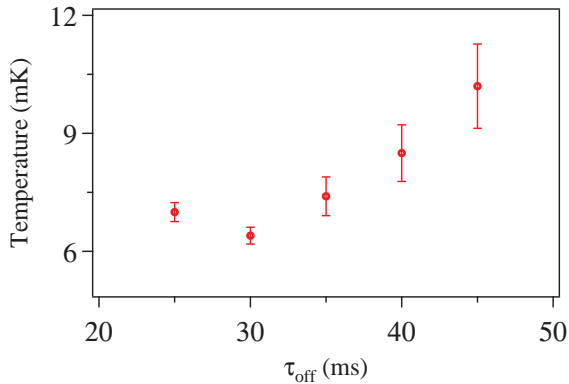}
\caption{(Color online)  Measured temperature of transferred cloud with two beams, as a function of $\tau_{\rm off}$ -- the time spent in the molasses. The larger error bars at higher values of $\tau_{\rm off}$ are due to loss of atoms, as explained in the text.}
\label{fig:Tvstau}
\end{figure}

The measured temperature as a function of $ \tau_{\rm off}$ is shown in Fig.\ \ref{fig:Tvstau}. As seen, the lowest temperature of 6.4 mK is reached with an off time of 30 ms. Beyond 45 ms, the atomic flux become very small and non-detectable, primarily because of collisional losses with background atoms This is the reason for the larger error bars at longer times. The lowest temperature of 6.4 mK is significantly higher than the temperature expected in the molasses, which is of order 100 \textmu K. This increase is due to heating in the blue-detuned pushing beams. Note that this is a transient process because the atoms move out of the pushing beams after some time. Seeing that the recoil velocity $\hbar k/m$ is 5.9 mm/s, the velocity of 28 m/s (corresponding to the peak velocity shown in Fig.\ \ref{fig:twobeamvel}) requires about 5000 absorption-emission cycles. This is a reasonable number given the $\sim$1 ms time spent within the pushing beams and the 26 ns lifetime of the excited state.

\section{Conclusions}

In summary, we have generated a cold pulsed beam of atoms by transferring atoms trapped in a 3D MOT. Atoms are loaded into the MOT from the low-velocity tail of thermal vapor generated by heating a getter source. The MOT is formed using a standard configuration of 3 retro-reflected laser beams and a quadrupole magnetic field. The atoms are further cooled in optical molasses by turning off the magnetic field of the MOT before the beams for transfer are turned on. Using two transferring beams inclined at $\pm \, 9 ^\circ$ to the vertical, the transferred cloud is seen to have a symmetric Gaussian distribution in velocity. This is in contrast to using a single beam for transfer, which results in a highly non-Gaussian distribution because the atoms continue to interact with the beam.

Though we have demonstrated a \textit{pulsed} atomic beam because we have to wait for the MOT to get loaded before the transfer, the scheme can be made \textit{continuous} if we use a 2D-MOT (instead of a 3D-MOT), as we have done with our recent work on Yb atoms in Ref.\ \cite{RSN13}. However, the disadvantage of this method is a much higher longitudinal temperature.

\section*{Acknowledgments}
This work was supported by the Department of Science and Technology, India. KDR acknowledges financial support from the Council of Scientific and Industrial Research, India. The authors thank S Raghuveer for help with the manuscript preparation.


\end{document}